# Radiobiological investigations at tumor cell lines by exploiting chrono-biological aspects of chronological dose distributions[+]

W. Ulmer

Radiotherapy of the Klinikum München-Pasing and MPI of Biophysical Chemistry, Goettingen

**Abstract**

Using 31P-NMR spectroscopy the chronological behavior of the ATP-metabolism of the tumor spheroids C3H-MA, 9L-Gliome and the mono-layer L1210 has been analyzed via decrease of the β-peak. All three cell lines show characteristic periods, and a homeostatic control cannot be recognized. Essential components of these periods are circadian (i.e. one day), circa-semiseptan (i.e. 3.5 days) and circa-septan (i.e. one week). The determination of the survival fractions provides an optimum exploitation of radiation damages, when the ATP-concentration assumes a maximum value. This optimum is reached, when all three cycles exhibit the ATP maximum, which is only possible by accounting for the circa-septan rhythm. The goal of this study is to elaborate an optimal fractionation scheme with regard to the irradiation of tumors.

[+]In memory of W. Duechting, who departed in December, 2010

## 1. Introduction

The diagnostic principle of traditional medicine tacitly assumes the validity of the so-called homeostatic regulation of physiological processes (e.g. blood pressure, norm values for various concentrations of hormones and enzymes found in blood), and with regard to therapy schemes similar assumptions are made by most widely neglecting some specific aspects of the time factor. Thus in some disciplines like radio-oncology it is only of interest whether a weekly dose should be better applied as one single high dose at once per week or given in daily fractions [1-2]. The specific biorhythms of patients often are ignored, which may be of relevance with regard to chronological dose distributions. This viewpoint includes the assumption that repair mechanisms and other biological processes obey homeostatic control.

Since the existence of characteristic biorhythms is proved in the wide range of biology starting with DNA/RNA-metabolism of single cells (microscopic level) up to macroscopic area of blood flow (blood pressure) through vessels and similar macroscopic phenomena in biology, it is obvious that the desired aim of an optimization of therapy schedules can only be reached by a further inclusion of chrono-biological aspects [3 – 4, 6 - 7 and references therein]. In this communication we shall particularly consider some aspects of the ATP metabolism as examples of interacting chemical systems with feed-sideward coupling besides nonlinear self-interaction [10 – 11]. The practical consequences of the biorhythms superimposed to the ATP metabolism are considered as examples of radio-oncology (irradiation of tumor spheroids of a mammary adeno-carcinoma of mice, 9L Glioma, L1210 (mono-layer), and determination of the dose-effect relationship after irradiation. Thus the principles of chrono-biology mainly discovered by Halberg et al [3 - 4] appear to be – besides theoretical insights – of great importance in practical medicine (optimization of therapy schedules in radio-oncology and pharmacology).

## 2. Material and methods

### 2.1. Growth curves

The spheroids considered in this communication and in the previously presented one [6, 8 – 9] can be adapted best by the fitting function for the time dependence of the cell number

$$N(t) = N_0 + (N_s - N_0) \cdot \tanh(t/T_0) \qquad (1)$$

$N_0$ refers to the initial number at t = 0 and $N_s$ to the number in the plateau phase. If $t = T_0$ in the growth phase, we obtain $\tanh(1) = 0.76$ and $t = 2T_0$ provides $\tanh(2) = 0.97$, i.e. the plateau phase has been reached. In the computer simulations of Duechting et al [8 – 9] a Gompertz function for the representation of the cell kinetics has been applied. The growth function (1) obeys a differential equation describing growth and inhibition (nonlinear term), which finally leads to a steady state:

$$dN/dt = A + B \cdot N - C \cdot N^2 \qquad (2)$$

The parameters A, B, and C are given by:

$$\left. \begin{array}{l} A = (N_s - 2 \cdot N_0)/(N_s \cdot T_0 \cdot (N_s - N_0)) \\ B = 2 \cdot N_0/(N_s \cdot T_0 \cdot (N_s - N_0)) \\ C = 1/(N_s \cdot T_0 \cdot (N_s - N_0)) \end{array} \right\} \qquad (2a)$$

The linear term $B \cdot N(t)$ is related to the usual exponential growth, which would be solely dominant for C = 0, whereas the nonlinear term $C \cdot N^2$ is connected to a cellular contact interaction and implies an inhibition of the spheroid growth. The nonlinear inhibition term cannot be interpreted as a cellular 'death'; the equilibrium state does not result from the transportation of necrotic cells from the inner domain of the spheroids via diffusion. This has never been observed in this way experimentally. The inhibitory term C might result from a specific cellular interaction evoked by the necrotic cells in the spheroid center. Similar conditions may also hold for the L1210 mono-layers. $T_0$ of the three cell cultures assumes the following order: $T_0$ (C3H-MA) = 10.07 days; $T_0$ (9L-Glioma) = 12.54 days; $T_0$ (L1210) = 8.68 days.

## 2.2. 31-P-NMR Fourier spectroscopy

Since the evaluation of the material referring to our investigations on spheroid growth, 31P-NMR-spectroscopy, and dose-effect-relationships in dependence of the chrono-biological rhythms have already been published [6 - 10], we shall only give a small summary.

The ATP-concentration has been determined by 31-NMR-Fourierspectroscopy with a Bruker spectrometer operating at 40.25 MHz. The internal reference system (δ-scale) of the 31P peaks was the resonance signal of PCr (phosphocreatine). Between PCr, ATP and ADP the following equilibrium holds:

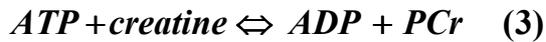

$$ATP + creatine \Leftrightarrow ADP + PCr \quad (3)$$

This relation is important with regard to many biological/biochemical processes in cells, since in all bio-energetic processes (muscles, biosynthesis of DNA, RNA, proteins and various other metabolic processes) much ATP in consumed, and *PCr* serves as a first pool of delivering ATP. With respect to irradiations ATP represents also a key molecule for repair processes of hit cells (e.g. Elkind repair, PLDR, repopulation, etc.). Dose-effect curves are significantly influenced by the ATP available in cells/tissue.

All measurements have been performed by **FID (free induction decay)** of the longitudinal relaxation time $\tau_1$. The quantitative, dose-dependent determination of the signal strength of the ß-ATP peak requires the use of the Feynman influence functional method to eliminate noise. In an earlier publication [6, 10] the dose dependence of the β-ATP peaks has been determined quantitatively, since its decrease represents a characteristic measurement of the ATP consummation. Each peak has been determined by a sequence of 200 pulses (repetition time of each pulse: 3 seconds). The longitudinal relaxation time $\tau$ is given by the FID sequences ($90^0 - \tau - 90^0$) according to the equation:

$$M_z(t) = (1 - \exp(-t/\tau)) \cdot M_0 \quad (4)$$

This equation reflects the decrease of the magnetic induction. In this communication we shall use the chronological behavior of ATP in the growth – and plateau phase to determine radiobiological dose-effect relations much more accurate. By accounting for chrono-biological parameters the possibility arises to gain further optimization aspects in radio-oncology.

The cell cultures (tumor spheroids C3H-MA and 9L-Glioma, and the mono-layers L1210, respectively) have been irradiated with 30 kV X-rays. The radiation exposition has been performed, when the spheroids had reached the plateau phase. The diameters of the spheroids in this phase amounts to ca. 0.8 - 1.9 mm. The colony-forming ability has been determined by clonogenic assays. The survival fraction S is normalized to 1, if no radiation dose is applied to the spheroid. The spheroids have been grown in the cytological laboratory of the MPI, Goettingen.

*2.3. Theoretical methods (tumor growth and survival functions)*

With regard to the measurement data in the next section we may restrict ourselves to define the survival function S(t) in the following way:

$$S(t) = N(t)/N_s \quad (5)$$

The time-dose dependence of S has been considered in detail previously [5] and may generally be represented by the power series:

$$\ln S = \sum_{n=1}^{\infty} a_n \cdot D^n \quad (6)$$

However, equation (6) requires the knowledge of an infinite number of coefficients $a_n$; therefore the restriction to simpler models is necessary. A very familiar approximation is the linear-quadratic model [2, 5], which accounts only terms up to second order:

$$\left. \begin{array}{l} -\ln S = \alpha \cdot D + \beta \cdot D^2 \\ S = \exp(-(\alpha \cdot D + \beta \cdot D^2)) \end{array} \right\} \quad (7)$$

This model has also been applied as an irradiation model in the papers of Düchting et al [8 - 9]. From equation (7) results that for very small doses D we obtain a usual exponential function with slope α, which is followed by a shoulder with increased slope (the order of magnitude depends on the cell line and radiation quality). If the applied doses still increase, the shoulder passes to an exponential function again, but with an increased slope. This domain cannot be represented precisely by the linear-quadratic model. Only by an overall fit that domain can be adapted, which is still in the environment of the shoulder. The measurement results to be presented in the next section can be adapted by equation (7) via least-square fit only, if a mean standard deviated of 25 % is permitted for the C3H-MA spheroid (the deviations of the other cell lines amount to ca. 35 % - 40 %). A very accurate adaptation can be reached, if the survival function S is represented by:

$$S = A \exp(-aD)/(1+B\exp(-(a+b)D)) \quad (8)$$

The parameter B has to satisfy B = A − 1, since the function S is normalized to 1 for D = 0. The mean standard deviation amounts to 2 % for the cell lines under consideration. Equation (8) exhibits the capability to represent S in the low dose domain as well as in the high dose domain by a corresponding exponential function. The formation/power expansion of *ln S* according to equation (8) provides the linear-quadratic model by restriction to terms of second order:

$$\left.\begin{array}{l} \alpha = (a+b)/A - b \\ \beta = (a+b)^2 \cdot \sum_{n=1}^{\infty} (A-1)^n \cdot (-1)^n \cdot n/2 \end{array}\right\} \quad (9)$$

Equation (8) is not only interesting for an accurate approximation method of dose-effect relations; its connection to the nonlinear equation (2) is also noteworthy. Convergence of the β-term is only reached, if A satisfies the condition: 1 < A < 2. By use of the relation (5) we are able to derive equation (8) by a nonlinear growth equation:

$$-N_S \cdot dS/dt = -dN/dt = -N_S (dS/dD) \cdot dD/dt = \lambda_1 \cdot N - \lambda_2 \cdot N^2 \quad (10)$$

The exponential decay due to irradiation is modified by the quadratic term $N^2$. If dD/dt is either constant or – compared to Elkind repair – rather short, then we can replace equation (10) by the simpler relation:

$$-dS/dD = a \cdot S - \rho \cdot S^2 \quad (11)$$

From this equation follows that repair processes in cells would be rigorously proportional to $S^2$ or $N^2$ even for extremely high doses D. This is however not true and if we assume that the repair capacity decreases by an exponential function, we obtain the final differential equation:

$$-dS/dD = a \cdot S - (a+b) \cdot B \cdot \exp(-b \cdot D) \cdot S^2 \quad (12)$$

The coefficient B = A − 1 has already been defined, and the solution of equation (12) is given by equation (8), which will be applied to Figures 4 – 6.

### 3. Results

The tumor spheroids C3H-MA, 9L-Glioma and the mono-layer L1210 belong to the rapidly growing tumors (in vitro and in vivo), and after ca. 20 days the plateau phase has been reached.

In Figures 1 - 3, the growth phase can clearly be verified by the peak height of 31P-NMR spectroscopy (ß-peak of ATP). In the plateau phase, the oscillations of ATP concentration can readily be registered. Since ATP is requested in the whole of all biochemical and energetic processes, and the total metabolism of ATP corresponds to the weight of the related living system, the signal height of the ATP concentration stands in a close relationship to the activity of the overall metabolism. A statistical analysis of the data shows that this activity underlines approximately rhythms of one day (circadian), 3.5 days (circa-semiseptan) and a week (circa-septan). In addition, a day periodicity is also existent. On this occasion it should also be mentioned that short periodic oscillations of the ATP concentration of the order of some minutes are already well-known in biological literature. It is evident, that the radio-sensibility should also depend on this periodicity. Halberg et al [4] have pointed out this connection many years ago. Therefore we may anticipate that the tumor cell lines are most sensitive with regard to dose-effect relations, when the ATP concentration assumes a maximum value. On the other hand, these relations should assume a less steep slope in the case of the minimal ATP concentration.

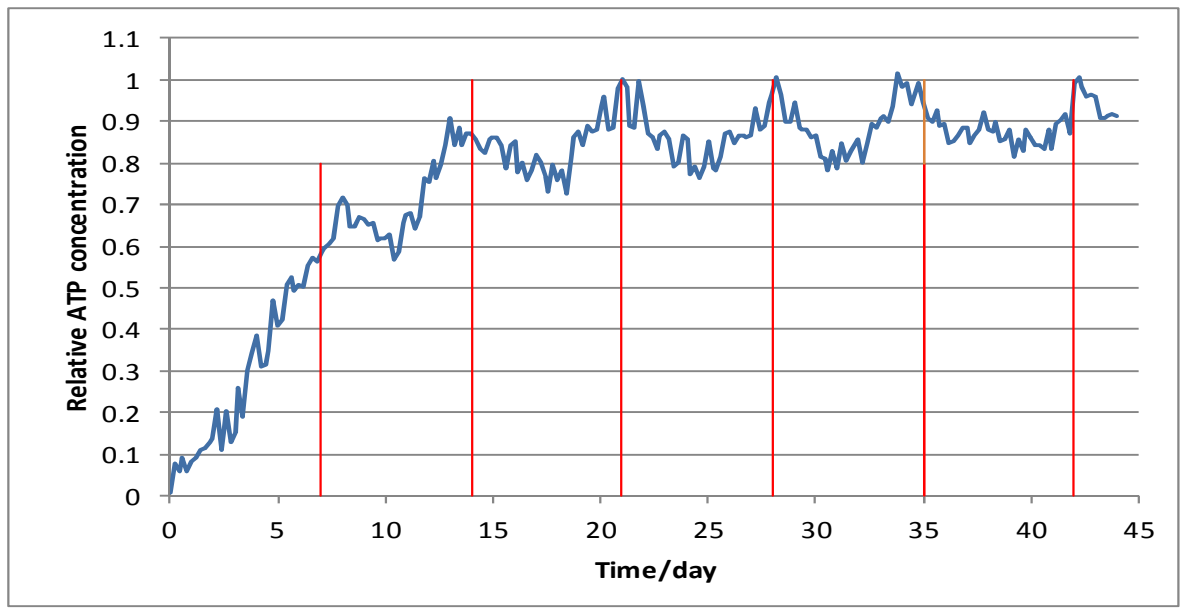

**Figure 1:** ATP concentration of the C3H-MA tumor spheroid in the growth – and plateau phase.

Thus the survival functions S(D) indeed exhibit this behavior for all three cases, which provide the possibility to exploit the maximal sensitivity of cells in radiation therapy. The adaptation of the survival curves by equation (8) works well, and the mean standard deviations do never exceed 2 %. The parameters a, b, and A according to this formula are listed in Table 1. Due to the errors of the linear-quadratic model beyond the shoulder we cannot use this model for the determination of the parameters of equation (8).

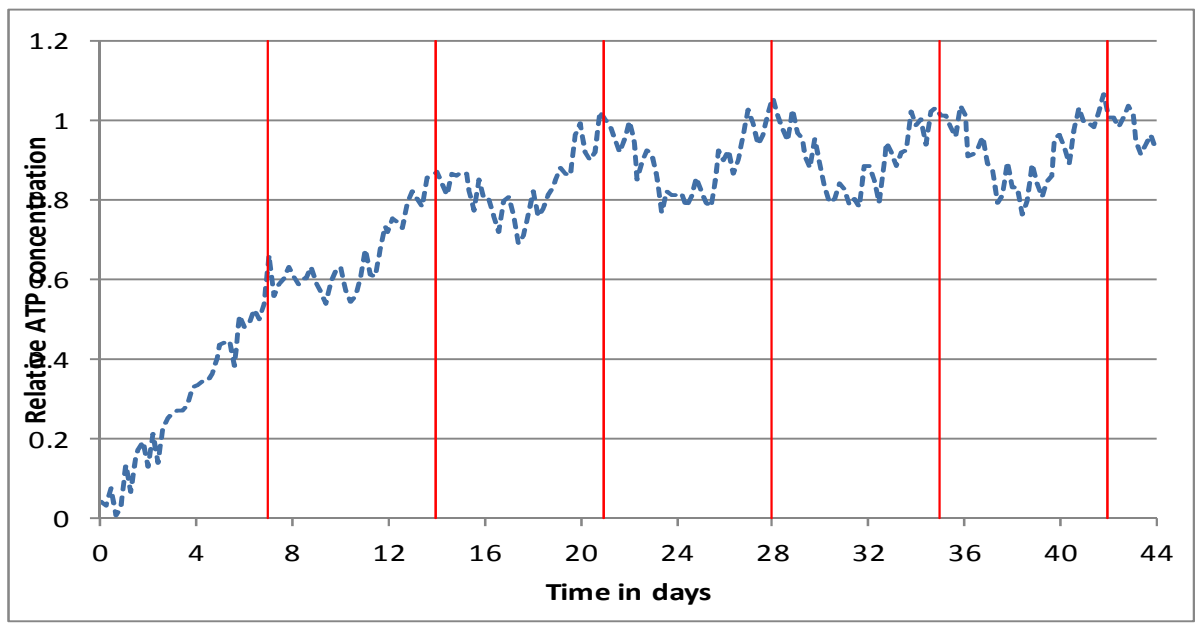

**Figure 2:** ATP concentration of L1210 cells (mono-layers) in the growth – and plateau phase.

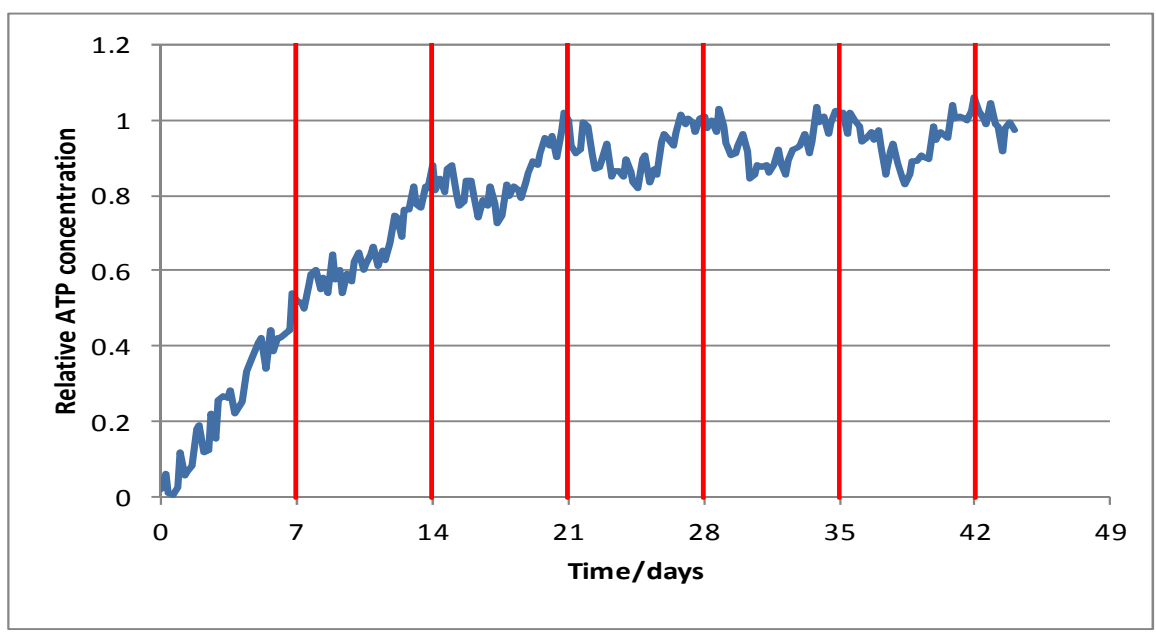

**Figure 3:** ATP concentration of 9L Glioma tumor spheroid in the growth – and plateau phase.

**Table 1:** Parameters a [in Gy$^{-1}$], b [in Gy$^{-1}$] and A of the survival function S according to equation (8).

| ATP maximum | a | b | A | ATP minimum | a | b | A |
|---|---|---|---|---|---|---|---|
| C3H-MA | 0.702 | 0.194 | 1.39 | | 0.599 | 0.106 | 1.59 |
| L1210 | 1.166 | 0.197 | 1.59 | | 0.907 | 0.198 | 1.96 |
| 9L-Glioma | 2.080 | 0.190 | 1.54 | | 0.867 | 0.186 | 1.76 |

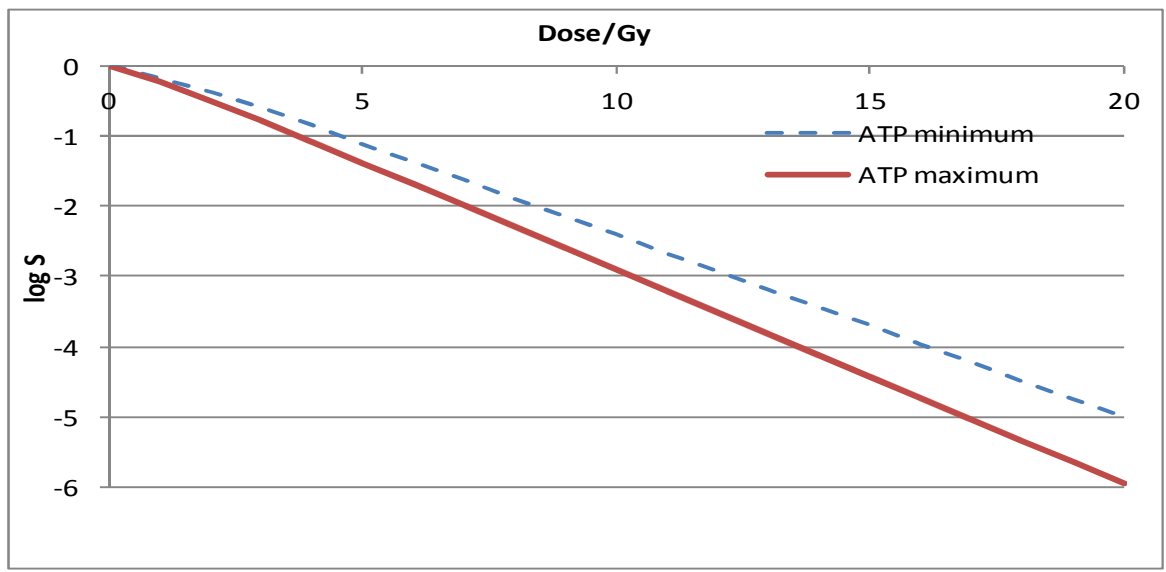

**Figure 4:** Dose-effect relations of C3H-MA tumor spheroid irradiated by 30 kV X-rays and determined at the ATP maximum (solid line) and minimum (dashed line).

A possible way to determine the parameters A, a, and b of equation (8) may work as follows: In a first step we use the tangent of the measured survival fraction (log S or ln S) in the high dose region to obtain A' and a' (both may not yet be the final values a, A). The parameter b' results from the initial slope at D = 0 according to equation (9). In a second step we use some further measured data inclusive the shoulder to correct the parameters of step 1 slightly by a least-square fit, which yields the final values a, b, A from a', b', and A'.

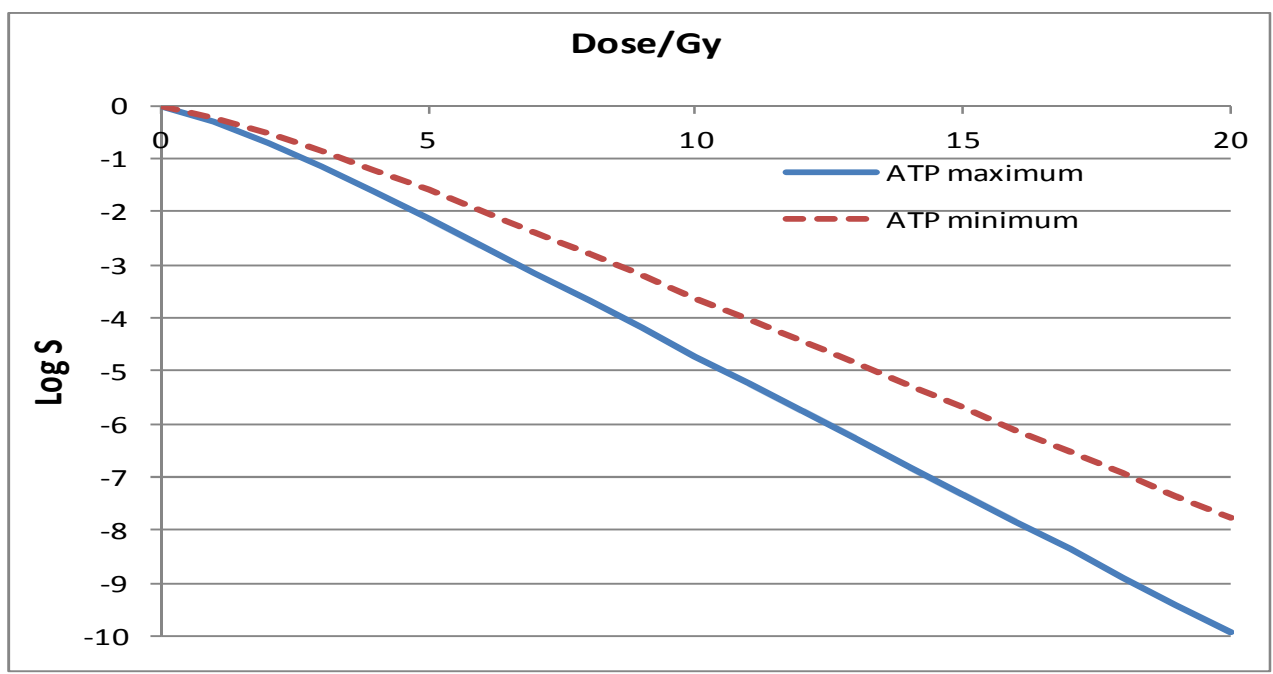

**Figure 5:** Dose-effect relations of L1210 tumor cells (mono-layers) irradiated by 30 kV X-rays and determined at the ATP maximum (solid line) and minimum (dashed line).

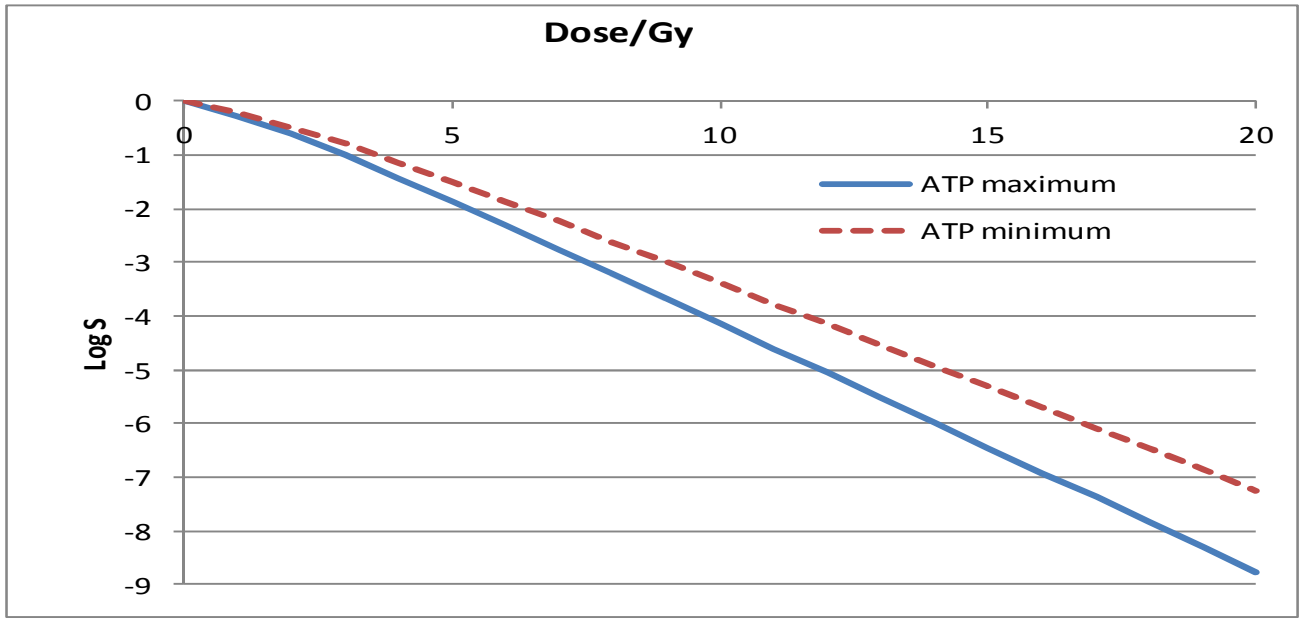

**Figure 6:** Dose-effect relations of 9L Glioma tumor spheroid irradiated by 30 kV X-rays and determined at the ATP maximum (solid line) and minimum (dashed line).

### 4. Conclusions

The homeostatic control in biology and medicine is rather insufficient and may only represent an average. The experimental data referring to ATP metabolism show the dominance of a circa-septan period. The fast oscillation component of ATP is well-known (about 20 minutes), but the more dominant week period exhibits the importance of chrono-biology in therapy (here: radio-oncology). It is due to the scientific work of Halberg et al [3 - 4] that the inclusion of chrono-biology in various therapy schedules offers the possibility of significant improvements of therapeutic results. The complete ATP metabolism is an example of a nonlinear reaction-diffusion system with feed sideward coupling to many other constituents in the citrate circle. Theoretical results seem to found that the basis of these nonlinear interaction systems (model aspects) in the presence of weak external magnetic fields provide wave-like processes and oscillations, which are a key in limit cycles discussed in evolution biology and chrono-biology [7, 10 - 11]. A practical consequence with relevance to radiotherapy results from the irradiation of the cell lines with ionizing radiation. The determination of the survival fraction S in dependence of the irradiated dose D confirms the ideas of Halberg et al [4]. There is a noteworthy difference (Figures 4 - 6), whether the cells have been irradiation in the ATP minimum or maximum. In the latter case the dose-effect function is much steeper. The results support the proposal of Wideröe [1] to abandon the standard fractionation. The dose-effect relations can be best adapted by the survival function (8). However, Wideröe only considered fractionation schemes with one single dose per week, and did not take into account chrono-biological aspects, which are certainly an inevitably necessary key for hypo-fractionation and therapy optimization. It appears that the delivery of high single doses in a therapy scheme makes only sense, if chrono-biological principles are accounted for. The research of Duechting et al [8 – 9], namely the computer simulation of cell growth and irradiation, has been extended by Stamatakos et al [12 – 13]. A successful extension should account for chrono-biological aspects.